\documentstyle[aps,epsfig,multicol]{revtex}
\begin{document}
\draft
\title { Nature of largest cluster size distribution\\
 at the percolation 
threshold}
\author{Parongama Sen}

\address{Department of Physics, University of Calcutta,
92 A.P. C. Road, Calcutta 700009, India.\\
e-mail paro@cucc.ernet.in}
\maketitle


\begin{abstract}
 Two distinct  distribution  functions $P_{sp}(m)$ and $P_{ns}(m)$ 
of the  scaled largest cluster sizes $m$ are obtained at the percolation 
threshold  by numerical simulations,  
depending on the condition  whether the lattice is actually spanned or not.
With $R(p_c)$ the spanning probability, the total  distribution    
of the largest cluster is given by
 $P_{tot}(m) = R(p_c)P_{sp}(m) + (1-R(p_c))P_{ns}(m)$.
The three distributions apparently
have similar forms in three and four dimensions while  in two
dimensions, $P_{tot}(m)$ does not  follow a familiar
form.  
 By studying  the first and second  cumulants of the 
distribution functions,   the different behaviour of $P_{tot}(m)$
in different dimensions may be quantified.

\end{abstract}

\begin{multicols}{2}
\pagebreak

Much has been investigated regarding the distribution of cluster 
sizes as far as percolating clusters are concerned \cite{hav,Ah,ps1}. 
There have been some
recent studies on the largest cluster size distribution below criticality 
and also distribution of smaller clusters at the percolation threshold 
\cite{ref1,ref2,martin,ps2}. The largest cluster size distributions, an example of
extreme value statistics, is relevant for several physical phenomena 
like fracture and breakdown.
One can obtain  distribution functions for the spanning cluster 
as well
as for smaller clusters
though the form of the distribution, especially that
of the percolating clusters is not simple.
In general, it is a non-Gaussian function.


Let $p$ be the probability that a site is occupied in a lattice. 
At the percolation threshold $p_c$, there   may or may not exist a spanning
cluster.  The spanning probability 
depends on many factors like the kind of percolation 
(site or bond), type of lattice, boundary conditions etc.
 \cite{Ziff}.
Spanning will occur with a certain probability less than one, and in the
study of spanning or percolating clusters at $p_c$,  
only those cases where the lattice spans are taken into account. 
On the other hand, for the largest cluster 
size distribution, the calculations will include 
all the configurations whether the lattice  spans or not. 
The largest cluster, in fact, enjoys a double role in the sense that 
it may or may not happen to be the spanning cluster. 
The probability of the smaller clusters being the spanning cluster 
is relatively much smaller.
 Even when the lattice is spanned, the largest cluster may  
or may not be the spanning one.  

The event of the lattice being spanned or not cannot be 
predicted a-priori in random percolation. 
 However, the consequence of the lattice being spanned or not 
would directly be reflected by the  nature of the distribution of the 
largest cluster which assumes different roles for the two cases.
We  find that the two distributions of the scaled mass or 
size $m$ of the largest cluster, denoted by 
$P_{sp}(m)$ and 
$P_{ns}(m)$ for the spanning case and the non-spanning case
respectively, 
are indeed different.
The scaled size $m = M/L^D$, 
where the  mass of the largest cluster  is $M$ in a lattice of 
linear dimension $L$ and $D$ is the   fractal dimension.  
The latter is related to the exponents of percolation and is
same for all clusters when they are ranked \cite{SAJ,ps2}.

The total distribution, which can be independently computed,
 is  actually the weighted sum of the two 
distributions:
\begin{equation}
P_{tot}(m) = R(p_c) P_{sp}(m)  +(1-R(p_c)) P_{ns}(m),
\end{equation} 
where $R_{p_c}$ is the spanning probability at $p_c$.
We investigate the nature of the 
distributions for hypercubic lattices in two, three and four dimensions, where
the Hoshen-Kopelman algorithm \cite{Hoshen} is used. 
Free boundary condition has been used in all dimensions, except that in two dimensions
we have also used helical boundary conditions for comparison.
In all cases, spanning has been considered from top to
bottom. The largest lattice sizes considered are $L=1600$,  $L=98$
 and $L= 27$ in two, three and four dimensions respectively with typically 
$10^6$  and $10^5$ configurations generated for the smallest  and largest 
lattice sizes. 

We first compare the distribution
of the largest and the spanning cluster  sizes in  percolating 
two-dimensional lattices (Fig. 1)  and find that they
are numerically indistinguishable almost always except for some cases
where the size of the clusters are very small.
One can ignore that difference and assume that at least for large values,
  the spanning cluster and the largest clusters are identical.
However, for consistency, we will
 consider the largest cluster in the spanning case strictly and not 
the spanning cluster if they are different.
This also takes care of the fact that we will avoid the ambiguity
arising due to the existence of more than one  
spanning cluster, which may happen in very few cases.  

Figures 2-4 show the three distributions for   
two to four dimensions with free boundary conditions.
For clarity, we have shown the distribution for a single  
lattice size in three and four dimensions, which represents the 
scaling  distribution.
Certain features  are clear from the figures: the total distribution
in two dimensions is quite different in shape compared to those in 
three and four dimensions. 
The distribution when the lattice is not spanning, $P_{ns}$, is much more 
sharply peaked in the higher dimensions while the width of  
$P_{sp}$, the distribution  
 for the spanning case, is of the same order 
 in different dimensions.
These features will be confirmed quantitatively later on.
The forms of the two distributions $P_{sp}$ and $P_{ns}$ are clearly 
different and it should be noted that
 it is not possible to get a collapse by any trivial scale transformation.

The  distribution function for the percolating 
clusters  has been fit to  the following forms \cite{hav,Ah,ps1}:  
a power law-exponential function
\begin{equation}
f(x)=ax^b\exp(-cx^d),
\end{equation}
or a double exponential function
\begin{equation}
f(x)=a \exp(-bx^{-c})\exp(-dx^e).
\end{equation}

What is important is the appearance of a number of  parameters 
in both the functional forms and estimates from
numerical results may  turn messy and involve large errors. 
The focus of the present study is,  however, not to 
obtain the precise  form of the  function but to compare the
gross features of the distributions. Even 
without a detailed study, it is obvious that the 
total distribution $P_{tot}$ in two dimensions has a form very
different from that of the individual distributions. 
For three or four dimensions, although  the  distributions
for the spanning and non-spanning cases are distinct, the total distribution
does not carry any signature that it was generated from these two.

That the behaviour of $P_{tot}$ in  two dimensions is different
becomes all the more apparent when one attempts to fit the  distributions
to familiar forms.  
We do this without emphasis on the accuracy of the
estimated  parameters. 
In two dimensions as well as in three and four,  the  two distributions
$P_{sp}$ and $P_{ns}$ 
 fit quite well to the 
form (3) with different values of the parameters
$a,b,c,d,e$. For example, in three dimensions
$c \sim 2.8$ and $e \sim  3.5$ for $P_{sp}$
and
$ c \sim  2.2 $ and $e \sim 1.5 $ for $P_{ns}.$
The total distribution, however, is of the same form only in three
and four dimensions. 
The values of the
exponents e.g.,  in three dimensions are 
$ c \sim 2.5$ and $ e \sim  1.5$ for $P_{tot}$.
Although in general the parameters are different,
we notice that among the   
exponents,  the values of  $c$  
are quite close for the three distributions  and   
$e$ is perhaps same for $P_{tot}$ and $P_{ns}$.  
In four dimensions also  the values of $c$ are 
comparable for the three distributions: $c$ lies between 1.9 and 2.0
but the values of $e$ are quite different. In two dimensions, 
 however, both $c$ and
$e$ are widely different for $P_{ns}$ and $P_{sp}$: 
$c \sim 2.5$ and $\sim 1.45$, 
 $e \sim 3.0$ and $\sim 10.8$ for the two respectively.

 Since the boundary condition plays an important 
role in percolation problems, we also evaluate the distributions 
with a different boundary condition, namely,  helical
boundary conditions, in two dimensions. The results are shown in 
Fig. 5. Here also, 
there are two  distinct distributions
$P_{sp}$ and $P_{ns}$. 
The cluster sizes will obviously be larger for helical boundary condition (HBC)
compared to the free boundary case (FBC). The shift in the probability 
distributions can be explained by this  but what is remarkable is that the
total distribution shows a hump on the left side, showing that
the form of the total distribution is again not conventional.
 The distribution, with the hump on the left side has a  less pronounced form of
the plateau-like region compared to the free boundary case.
 Such a hump has also been observed for small sizes using a renormalisation 
group scheme \cite{martinpc} and periodic boundary
conditions. Hence one can conclude that in general
the total distribution for the largest cluster size is not in a 
familiar form, the effect being strongest in open boundary case. 

In order to understand the difference in the behaviour of the
total distribution in different dimensions we note the following points. 
The reason for the characteristic structure 
of $P_{tot}$ in two 
dimensions must be traced back to the features of the
two independent distributions from which $P_{tot}$ is generated.
$P_{tot}$ typically has a plateau like region (FBC)
or a weak two peaked structure (HBC).
This  could be due to
two reasons: either $P_{sp}$ and $P_{ns}$, which are both peaked,
have negligible overlap; or the width of the distributions are
comparable. As the distributions are normalised,
this would imply the heights of the peaks are 
comparable. The value of $R(p_c)$ should not 
play any part as it is not particularly different in different
dimensions. 

The overlap between the distributions are
not negligible as one can observe from  the figures. In order to 
compare the behaviour in two, three and four dimensions, we
evaluate the ratios $r_1 =  
m_{sp}^{(1)}/m_{ns}^{(1)}$ 
and $r_2 =
m_{sp}^{(2)}/m_{ns}^{(2)}$ 
where $m^{(r)}$ is the $r$th cumulant of the distributions.
The subscripts on $m$ denote the spanning and non-spanning
events as usual.
The first measure, if high,  will indicate  that the  peaks of the
distribution are far apart as $r_1$  is a measure of the
mean scaled cluster size and lies close to the peak. 
The second ratio is roughly a ratio of the widths of the distributions
which in turn is an estimate of the ratio of the height of the peaks.

We notice very interesting behaviour of the two ratios 
 defined above and shown in Fig. 6.
The first ratio $r_1$ varies between  1.6 and 1.7 with some weak
dependence on dimensionality.
$r_2$ on the other hand, shows  
 strong  dimensional dependence. 
The values  of $r_1$ and $r_2$ in two dimensions with   HBC
and FBC  
indicate that  both depend on boundary conditions.
In two dimensions, $r_2$  is close to 1, indicating that
the peaks of the distributions lie at comparable heights. In
higher dimensions, $r_2$ increases  and differs significantly  from unity.
The latter may therefore be exclusively responsible
for the different behaviour of the total distributions in different dimensions.

The observation 
that there exist  two separate distributions for the largest cluster sizes may 
not appear very surprising.
However, it has not been noted in any previous study although distribution
functions for percolation is a much studied problem.
 Also, usually in percolation,
quantities and their distributions  have the same qualitative behaviour in 
dimensions
below the upper critical dimension. The spanning cluster distributions
for example, in different dimensions, could be fit to the same form \cite{ps1}.
The smaller clusters had  distribution functions
which had familiar form, very small clusters following
a Gaussian distribution presumably in two dimensions \cite{ps2}. 
This is perhaps the first
time a novel feature is  seen to be present in two dimensions and absent
in higher dimensions as far as the total distribution for largest clusters are 
concerned. Even if one argues that  
 $P_{sp}$ and $P_{ns}$ should be considered as  the   
more fundamental 
distributions, 
rough estimates of the 
exponents for the distributions for these two 
showed that in two dimensions, they are different by a much larger
margin than in  higher dimensions, indicating that the 
two dimensional case is markedly different regarding distributions of largest
cluster sizes. 
This is an effect independent of boundary conditions although
the effect may vary for different cases.

Since the largest and the spanning clusters coincide in most cases,
$P_{sp}$ may be regarded as a previously known distribution. Hence $P_{ns}$
is the new distribution obtained from the study. 
We have made a brief comparison
of $P_{ns}$ with the distribution of largest cluster sizes
below $p_c$, as in both cases the lattice does not span. The
cumulative distribution $Q_{cum}(x)$ of the distribution $Q(m_l)$ of the 
largest cluster size $m_l$ below $p_c$ is well-known and has
the form \cite {ref1,ref2,martin}
\begin{equation}
Q_{cum}(x) =  Q(m_l<x) \sim \exp(-\exp(-\lambda_1 x + \lambda_2)) 
\end{equation}
such that $\ln(-\ln(Q_{cum}(x))$ is a straightline when plotted versus $x$.
We recover this behaviour but find that for $P_{ns}$, the
behaviour of the corresponding cumulative 
distribution is much more complicated.
Hence one can conclude that $P_{ns}$ is a completely independent
distribution and different from any previously known distribution in percolation.

We make one last remark: the existence of two distinct distributions at 
$p_c$  is
 not a unique feature of the largest cluster only,
it is true for any  cluster which does not span the 
lattice. But the total distribution
will behave differently depending on the rank of the cluster.

\medskip

I thank Martin Bazant for the numerous discussions
which inspired the present work and also for sending unpublished 
results. I am also grateful to Dietrich Stauffer and an anonymous referee
for comments. The computations were done on a Origin 200 at the Calcutta 
University Computer Center.

\end{multicols}
 

\begin{figure}
\psfig {file=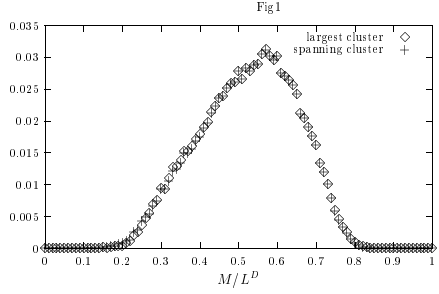,width=10cm}
\vskip -3in
\caption{The comparison of the distributions for the largest cluster and
the spanning cluster sizes are shown in two dimensions for $L= 600$ with
open boundary conditions.}
\end{figure}
\begin{figure}
\psfig {file=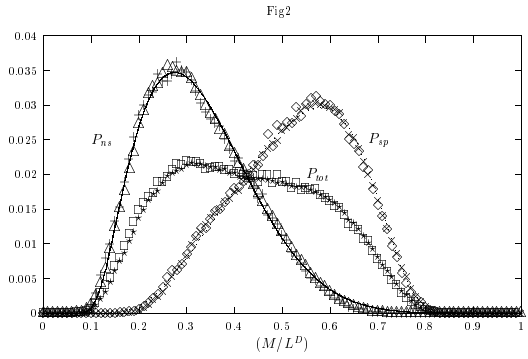,width=10cm}
\vskip -2.5in
\caption{The three distributions $P_{sp}$, $P_{ns}$ and $P_{tot}$ for the
scaled mass in two dimensions are shown for two different system sizes $L =
400$ (represented by $\times$, $\triangle$,  and $\star$  respectively) 
and $L=1000$. 
The best fit lines of the form (3)
are also shown for $P_{sp}$ and $P_{ns}$.}
\end{figure}
\begin{figure}
\psfig {file=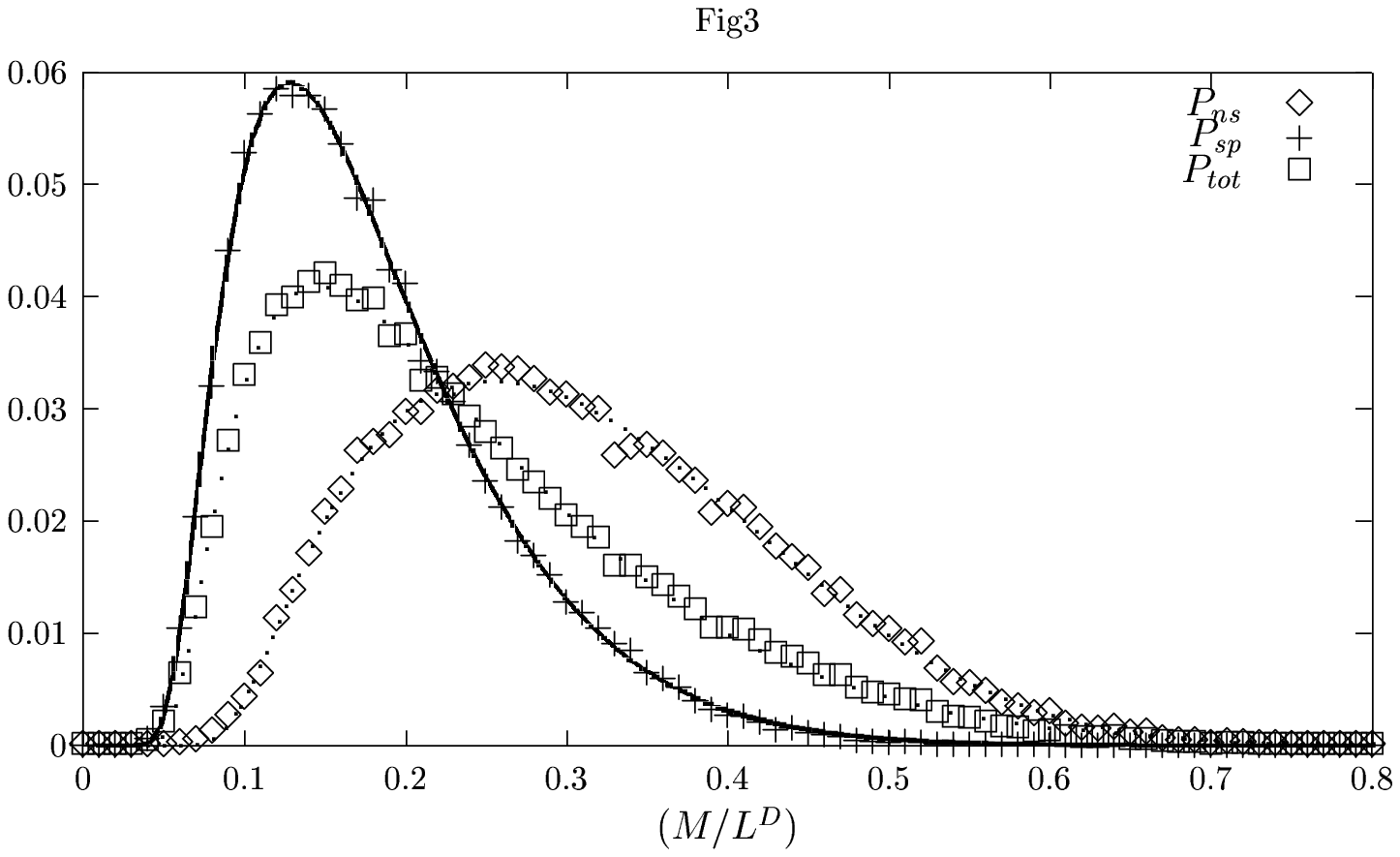,width=9cm}
\vskip -2in
\caption{The three distributions $P_{sp}$, $P_{ns}$ and $P_{tot}$ for the
scaled mass in three dimensions are shown for  $L =
60$  along with the  best fit lines of the form (3).}
\end{figure}
\begin{figure}
\psfig {file=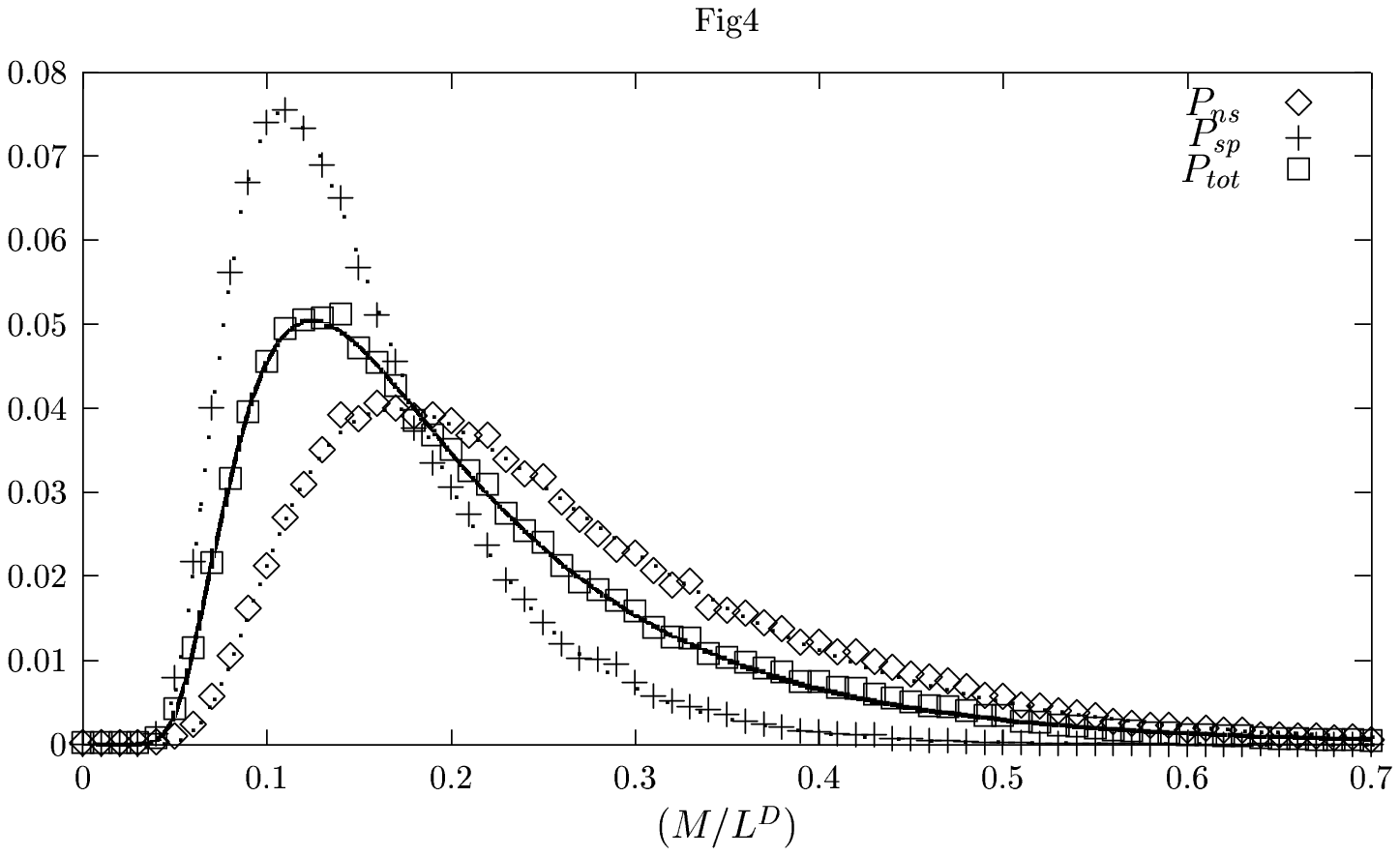,width=9cm}
\vskip -2in
\caption{The three distributions $P_{sp}$, $P_{ns}$ and $P_{tot}$ for the
scaled mass in four dimensions are shown for  $L =
21$ alongwith the  best fit lines of the form (3).}
\end{figure}
\begin{figure}
\psfig {file=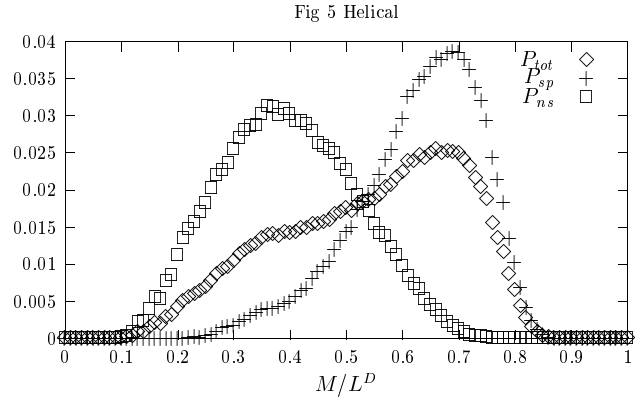,width=15cm}
\vskip -3in
\caption{The three distributions $P_{sp}$, $P_{ns}$ and $P_{tot}$ for the
scaled mass in two  dimensions are shown for  $L =
600$ with helical boundary conditions.}
\end{figure}
\begin{figure}
\psfig {file=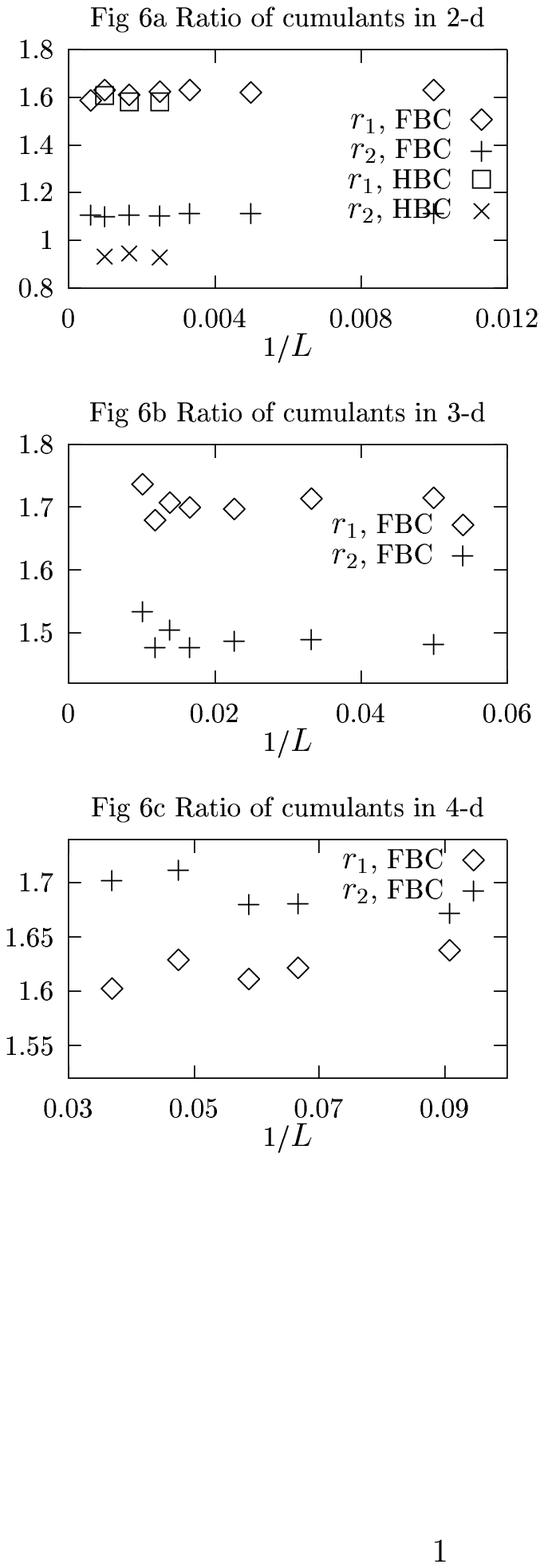,width=16cm}
\vskip -3in
\caption{The ratios $r_1$ and $r_2$ defined in the text are shown  in
two, three and four dimensions (FBC and HBC indicate free and
helical boundary conditions respectively).}
\end{figure}


\begin{references}
\bibitem{hav}
S. Havlin, B. L. Trus, G. H. Weiss and D. Ben-Avraham,  J. Phys A {\bf 18} L247 (1985). 
A. U. Neumann and S. Havlin J.  Stat.  Phys.  {\bf 52} 203 (1988)  
 H. Saleur and B. Derrida, J. Physique {\bf 46} 1043 (1985).  
\bibitem{Ah} J.-P. Hovi and A. Aharony,  Phys. Rev. E  {\bf 56} 172 (1997).
\bibitem{ps1} P. Sen,  Int.J. Mod. Phys. C {\bf 10 } 747 (1999)   
\bibitem{ref1} P. M. Duxbury and P. L. Leath. J. Phys. A {\bf 20} L411 (1987).
\bibitem{ref2} M. I. Zeifman and D. Ingman, J. App. Phys.  {\bf 88} 76 (1987).
\bibitem{martin} M. Bazant, Phys. Rev E {\bf 62} 1660 (2000). 
\bibitem{ps2}  P. Sen, J. Phys. A {\bf 32} 1623 (1999).
\bibitem{Ziff} R. M. Ziff, Phys. Rev. Lett. {\bf 69} 2670 (1992). 
\bibitem{SAJ}
N. Jan,  D. Stauffer and A. Aharony, J. Stat. Phys. {\bf 92} 325 (1998).
\bibitem{Hoshen} J. Hoshen and R. Kopelman, Phys. Rev. B {\bf 14} 3428 (1976). 
\bibitem{martinpc} M. Bazant, unpublished. 
\end{references}
\end{document}